\documentclass[]{aa}
\usepackage{times,pictex,graphicx}

\begin{document}

\title{The Evolution and Simulation of the Outburst from XZ~Tauri \\ - A Possible 
EXor?}

\author{Deirdre Coffey \inst{1}, Turlough P. Downes \inst{2}
\and Thomas P. Ray \inst{1}}

\offprints{D. Coffey, dac@cp.dias.ie}

\institute{School of Cosmic Physics, Dublin Institute for Advanced Studies, 
5 Merrion Square, Dublin 2, Ireland
\and Dublin City University, Dublin 9, Ireland}
\date{Received date ;accepted date}
\authorrunning{Coffey et al.}
\titlerunning{The Evolution and Simulation of the XZ~Tauri Outburst}

\abstract{
We report on multi-epoch HST/WFPC2 images of the XZ~Tauri binary, and its
outflow, covering the period from 1995 to 2001. Data from 1995 to 1998
have already been published in the literature. Additional images, from
1999, 2000 and 2001 are presented here. These reveal not only further
dynamical and morphological evolution of the XZ~Tauri outflow but also
that the suspected outflow source, XZ~Tauri North has flared in EXor-type
fashion. In particular our proper motion studies suggests that the recently
discovered bubble-like shock, driven by the the XZ~Tauri outflow, is slowing
down (its tangential velocity decreasing from 146\,km\,s$^{-1}$ to
117\,km\,s$^{-1}$). We also present simulations of the outflow itself, with
plausible ambient and outflow parameters, that appear to reproduce not
only the dynamical evolution of the flow, but also its shape and emission
line luminosity. 

}

\maketitle


\section{Introduction}

XZ~Tau is a classical T-Tauri binary system with a separation of 0$\farcs$3 
(\cite{Haas90}) and is located in the well known Lynds 1551 
star-forming region some 140\,pc away (\cite{Elias78}). The system was first 
found to have an associated Herbig-Haro (HH) outflow through ground-based CCD 
imaging and spectroscopy (\cite{Mundt88}; \cite{Mundt90}). These early 
observations revealed 
a bipolar optical 
flow that could be traced to at least 10$\arcsec$ on either side of the binary 
at 
a position angle of 15\,$\degr$. The first Hubble Space Telescope (HST) Wide 
Field Planetary Camera 2 (WFPC2) images of XZ~Tauri taken in 1995 show a 
bubble of emission nebulosity extending 4$\arcsec$ to the north of the system 
(\cite{Krist97}, hereafter K97).  Further images, taken 3 years later, 
show dramatic structural changes as the bubble expanded and altered from being 
centre-brightened to limb-brightened, suggesting the formation of a HH bowshock 
(\cite{Krist99}, hereafter K99).

Ground-based photometry of XZ~Tau from 1962 to 1981 (\cite{Herbst94}) has 
shown variations, of almost two magnitudes in the V band, for the binary as 
a whole. Such variations are common amongst young stellar objects (YSOs). 
Of the two components, the southern one has been, at least until recently, 
optically brighter and thought to be the 
more evolved star. Its companion, however, dominates at infra-red wavelengths 
and is probably of higher overall luminosity (\cite{Haas90}). Recent Faint 
Object Spectrograph (FOS) observations, however, unexpectedly found the 
northern component to be optically brighter (\cite{White01}, hereafter WG01), 
a result that we will discuss further in the light of our findings. For this 
system therefore it seems more appropriate not to use the terms primary 
and secondary but instead we will adopt the nomenclature, used in K97, of XZ 
Tau North and South. 

We report here analysis of further HST Archive WFPC2 images of the XZ~Tau 
system and outflow from 1999, 2000 and 2001. These data show not only ongoing 
changes in the outflow but a dramatic brightening of XZ~Tau North in the 
optical suggesting that it may be an EXor. We also simulate the outflow in an
attempt to reproduce its dynamical and morphological evolution.
    

\section{Data}

High resolution archival WFPC2 images of XZ~Tauri were obtained for 3 
consecutive epochs:  1999 February 3; 2000 February 6; and 2001 February 10. 
The binary was at the same approximate location on the Planetary Camera 
(spatial sampling = 0$\farcs$04555\,pixel$^{-1}$) for all frames.  
Four filters were used:  H$\alpha$ (F656N); [SII] (F673N); 
R-band (F675W) and I-band (F814W).  The frequency and duration of the short 
exposures for each filter were: 
1$\times$120\,s; 1$\times$180\,s; 2$\times$6\,s; and 2$\times$6\,s 
respectively. Two long exposures for each filter of 1000\,s were 
also extracted from the HST Archive, all of which were saturated at the 
location of XZ~Tauri.  No short exposures 
were made for the I-band filter in the final year.  

Previously published WFPC2 archival data (K99) for XZ~Tauri facilitated proper 
motion measurements. These data comprised images from 1995 January 5 and 1998 
March 25.  The 1995 images used were 2$\times$600\,s exposures in the R-band 
and 1$\times$3\,s exposure in the I-band.  The latter allowed us to determine 
the stellar positions, as no R-band short exposures were taken and the long 
exposures were saturated in the vicinity of the star. The 1998 images were 
taken in the R-band only:  2$\times$6\,s and 2$\times$1100\,s exposures.  All 
frames were processed 
through the standard HST pipeline and each set of double exposures was combined 
to eliminate cosmic 
rays.

The accumulated data yielded high resolution images of XZ~Tau spanning a 
total of 6 years.  We analyse these data in Section 3, under the two headings: 
Outflow Structure, Proper Motions $\&$ [SII] Luminosity and Stellar Astrometry 
$\&$ Photometry.


\section{Results}


\subsection{Outflow Structure, Proper Motions $\&$ [SII] Luminosity}
\label{structure_etc}

\begin{figure*}
\resizebox{\hsize}{!}{\includegraphics[scale=0.77]{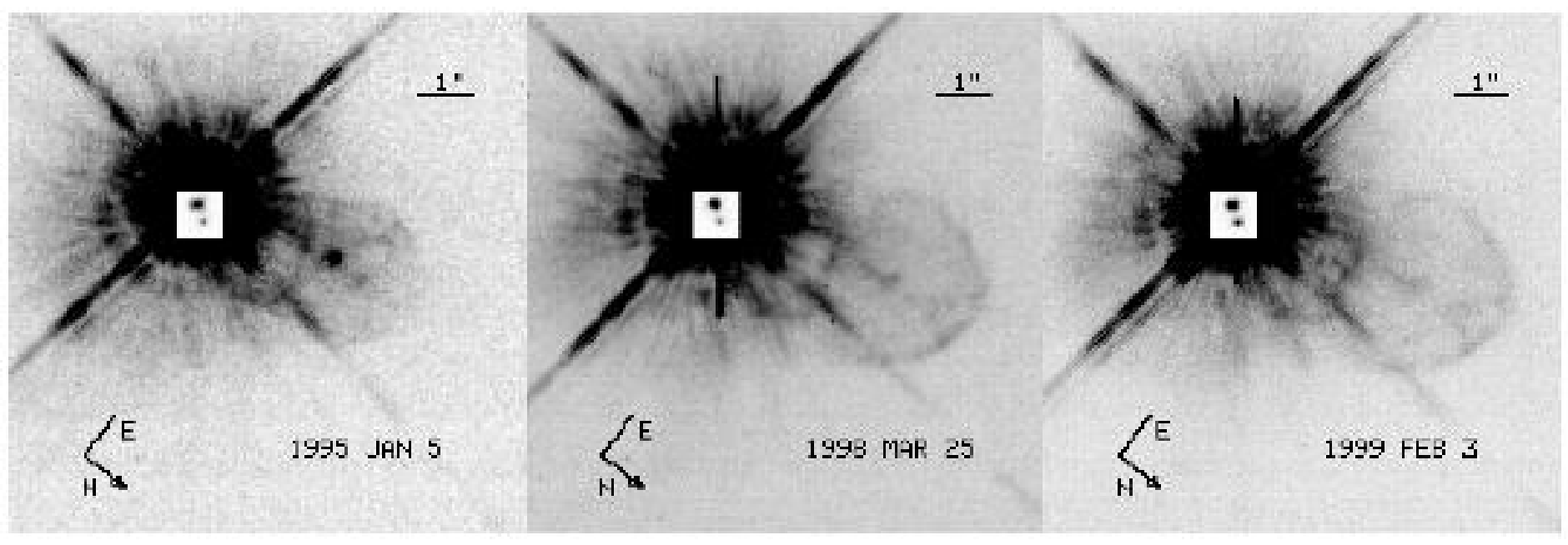}}\\
{\includegraphics[scale=0.63]{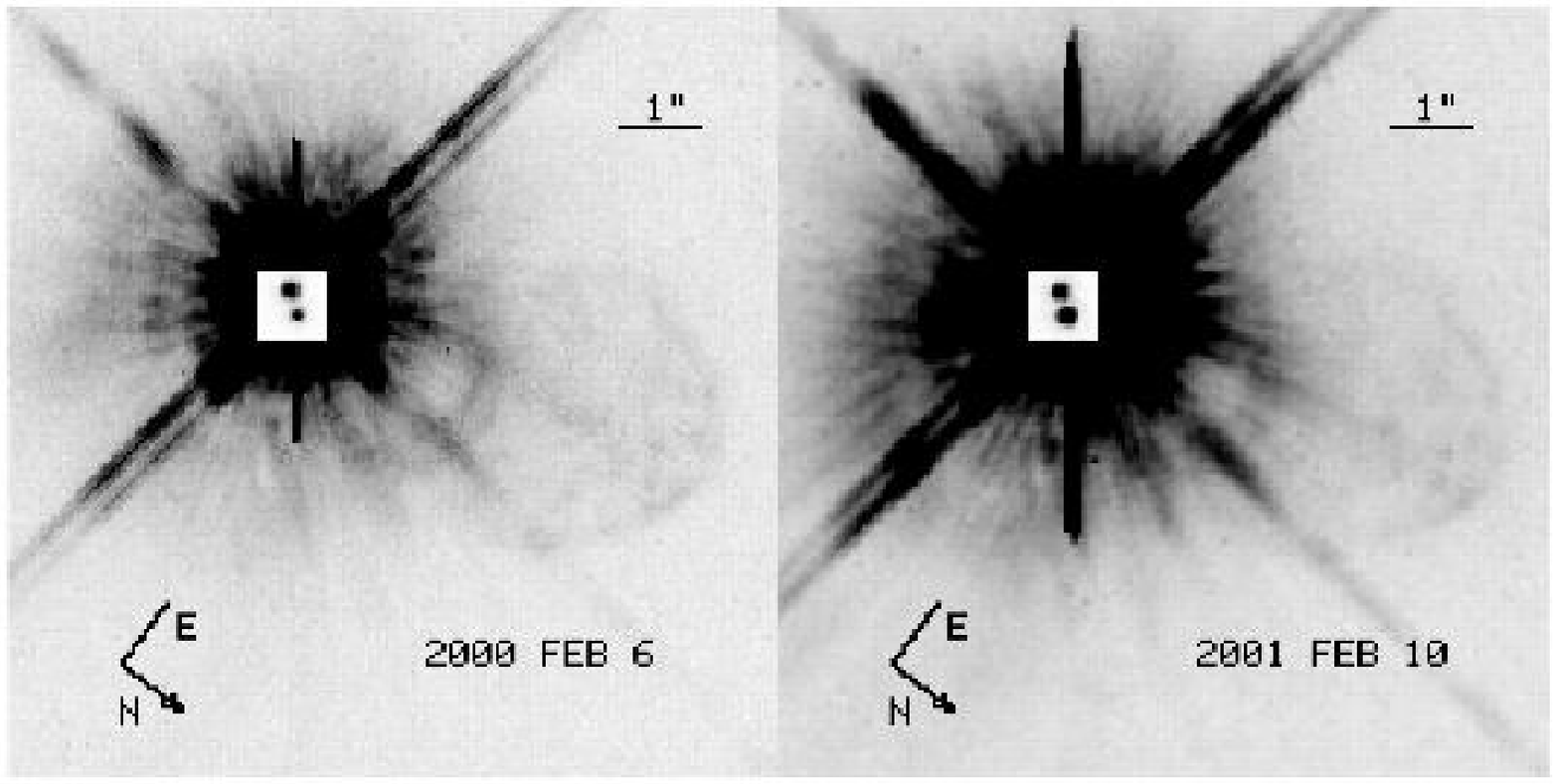}}
{\includegraphics[scale=0.5]{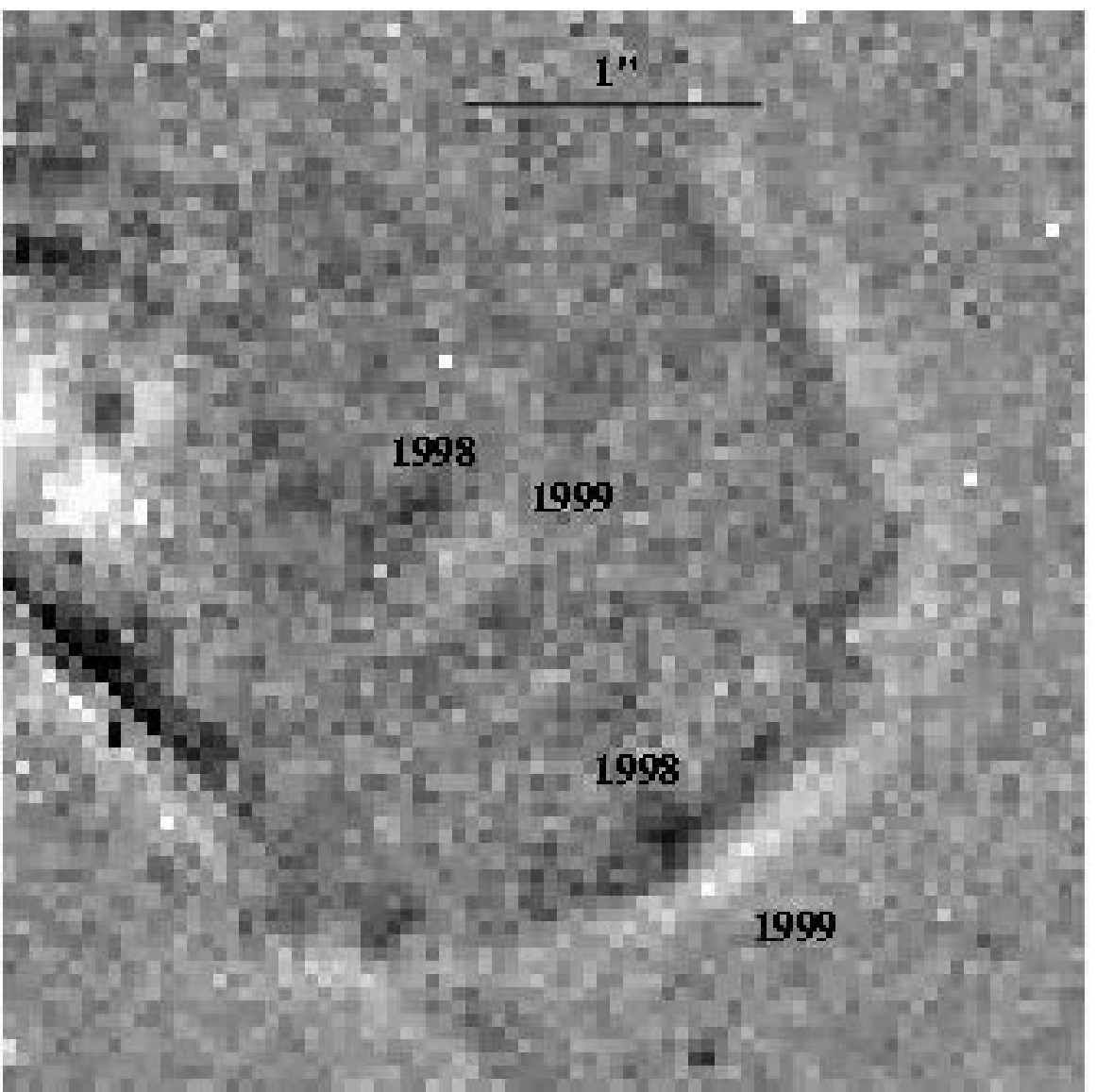}}
\small{\caption{HST/WFPC2 F675W (R-band) images of the XZ~Tau outflow on: 1995 
Jan 5; 1998 Mar 25; 1999 Feb 3; 2000 Feb 6; and 2001 Feb 10. The short exposure 
image of the 
binary is superimposed on the 
long exposure image in each case. The bottom right frame is an R-band 
difference image 1999-1998 showing proper motion in the jet and the outer 
shock.  
\label{xztaupm}}}
\end{figure*}

The Planetary Camera images for each year were aligned and their orientation 
corrected where necessary. A montage of R-band images, covering the full 
timespan, is shown in Fig.\ \ref{xztaupm}. As previously observed (K99), there 
appears
to be two edge-brightened ``bubbles'', i.e. inner and outer shocks, similar to 
those seen with HST close to DG Tau (\cite{Bacciotti00}). By 2001, the rapidly 
fading outer shock reached a distance of approximately 800\,AU from the binary, 
while the inner shock has travelled approximately half as far. 

Although remarkably bright in 1995, the knots in the XZ~Tau jet appear 
very faint by 2001. The jet's PA is around 15\,\degr\ in line
with the major axis of the elongated outer bubble and the known extended 
optical outflow. The knots however are not in a perfectly straight line as some 
deviation is evident and reminiscent of similar behaviour seen in flows like  
the HH\,34 jet close to its source (\cite{Reipurth02}). 

Significant proper motions were observed as the bubble expanded away from 
the XZ~Tau. Tangential velocities for the central jet knot and outer shock
(see for example the bottom right ``difference'' frame in Figure 
\ref{xztaupm}) were obtained from the long F675W exposures. The latter 
were used since they include the main HH emission lines, i.e.\ 
[OI]$\lambda\lambda$6300,6363, H$\alpha$ and [SII]$\lambda\lambda$6717,6731. 
The distance to the binary was assumed to be 140\,pc as in K99. In Table\ 
\ref{ProperMotions} we list the derived tangential velocities as a function of
epoch for the outer shock and an average velocity for the jet. Longitudinal sizes and 
speeds were calculated along the jet axis (P.A.$\sim$15\,\degr\ ). Transverse 
sizes and speeds were calculated perpendicular to the jet axis at the bubble's widest point, 
(specifically, the point where the source-to-shock line forms an angle of 20\,\degr\ with the 
jet axis in the 1995 image). As the shock expands it grows fainter and so its width is not 
given for later observations. The central knot also grows very faint, especially at later 
epochs, so we 
prefer to quote an average for the jet speed.  Finally, although the inner 
shock or bubble 
was seen to expand, changes from year to year proved difficult to trace and so 
no tangential 
velocity is given here.  

\begin{table} 
\begin{center}
\scriptsize{\begin{tabular}{lcccc}
\hline\hline
Observation			&Distance 	&Bubble 	&Longitudinal 		&Transverse		\\
Date/Interval			&from source	&Width		&Speed			&Speed			\\
				&(AU)		&(AU)		&(km\,s$^{\rm -1}$)	&(km\,s$^{\rm -1}$) 	\\ \hline 
1995 Jan 5			&598~$\pm 16$	&336~$\pm 25$	&-			&-			\\ 
1995 Jan 5 - 1998 Mar 25 	&697~$\pm 10$	&405~$\pm 14$	&146~$\pm 28$ 		&101~$\pm 42$		\\ 
1998 Mar 25 - 1999 Feb 3	&721~$\pm 9$	&423~$\pm 5$	&130~$\pm 73$ 		&105~$\pm 83$		\\ 
1999 Feb 3 - 2000 Feb 6		&746~$\pm 11$	&-		&119~$\pm 65$ 		&-			\\ 
2000 Feb 6 - 2001 Feb 10	&771~$\pm 10$	&-		&117~$\pm 69$ 		&-			\\ 
				&		&		&			&			\\ 
1998 Mar 25 - 2001 Feb 10	&-		&-		&130~$\pm 84$ (jet)	&-			\\ \hline
\end{tabular}}
\end{center}
\small{\caption{Projected sizes and speeds of the XZ~Tauri outer bowshock (and 
an average speed of the 
jet knot) as it 
evolves. Distances are quoted 
for the latter date of the observation interval. Errors are at the three sigma 
level. 
\label{ProperMotions}}}
\end{table}

From Table\ \ref{ProperMotions} it is clear that we see a deceleration of the 
outer shock. 
Although the errors appear large, if we consider the interval 1998-2001 we 
obtain a 
longitudinal speed of 121\,($\pm$~24)\,km\,s$^{\rm -1}$ (cf. 
146\,($\pm$~28)\,km\,s$^{\rm -1}$ from 1995 to 1998) suggesting the shock front is decelerating. Simulations were carried out in order to model the observations using plausible jet and ambient medium parameters (see $\S$3.3 below). Observerd speeds are as expected in order to maintain the observed low aspect ratio of the bubble, and are in the same region as those given by the simulation, (see Table\ \ref{sim_proper_motions} below). As a further check the simulated and observed [SII]$\lambda\lambda$6317,6731 luminosities were compared. The latter was found from the 2001 F631N images to be 2.5\,($\pm$\,3.6)\,$\times$$10^{\rm 28}$\,erg\,s$^{\rm -1}$, excluding the inner shock region which is contaminated with diffraction spikes and nebulosity near the star. If we assume an extinction of A$_v$$\sim$1.39 towards the bubble, i.e. the same as towards the binary (WG01), the intrinsic luminosity increases to $\sim$4.3\,$\times$10$^{28}$\,erg\,s$^{-1}$. 


\subsection{Stellar Astrometry $\&$ Photometry}

Using the short R band exposures, the separation and position angle of the 
binary were determined for the different epochs and are listed in Table\ 
\ref{BinaryStar}. Within errors, no changes were detected in the separation
of the XZ~Tau binary although an average decrease of 0.75\,\degr\,yr$^{-1}$ 
was observed in its position angle, a value somewhat higher than derived by 
K99 based on earlier HST data alone (of 0.5\,\degr\,yr$^{-1}$) but at the same 
time lower than that
found by \cite{Woitas01} using speckle interferometry (of $\sim$1.3\,\degr\,yr$^{-1}$). 
Following K99, if we assume a face-on 
orbit, the total mass for the binary is about 0.3\,M$_{\odot}$. Although such 
a value is clearly too low, note that the combined mass is very sensitive to 
the projection angle. Such a figure is certainly lower than the estimates 
of \cite{HartKen03} and WG01 suggesting a combined 
mass closer to 1\,M$_{\odot}$. 

\begin{table} 
\begin{center}
\scriptsize{\begin{tabular}{lllll}
\hline\hline
Date of 	&Separation 	&Position Angle		\\
Observation	&(arcsec)	&(degrees)	 	\\ \hline

1995 Jan 5	&0.302   	&146.35~$\pm 0.38$		\\
1998 Mar 25	&0.299   	&145.50~$\pm 0.25$		\\ 
1999 Feb 3	&0.303   	&144.85~$\pm 0.26$		\\ 
2000 Feb 6	&0.300   	&143.30~$\pm 0.25$		\\ 
2001 Feb 10	&0.298   	&142.38~$\pm 0.24$		\\ \hline

\end{tabular}\\}
\end{center}
\small{\caption{Separation and position angle of the XZ~Tau binary. 
Errors in separation angles are estimated to be~$\pm 0.005$.
\label{BinaryStar}}}
\end{table}

\begin{figure}
\begin{center}
{\includegraphics[scale=0.6,angle=-90]{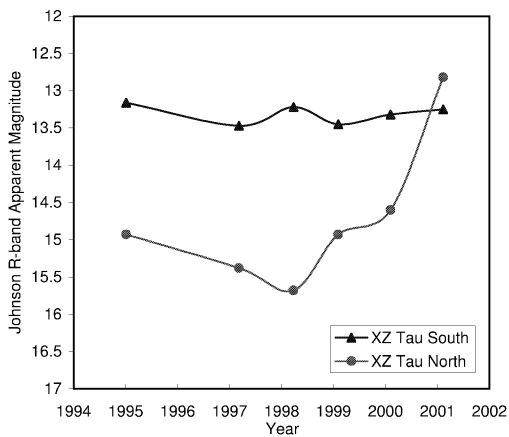}}
\end{center}
\small{\caption{Stellar Johnson R-band apparent magnitudes of the 
XZ~Tauri binary. 
\label{stellarmagnitudes}}}
\end{figure}

Broadband R and I magnitudes were calculated for the binary, using the method 
outlined in the 
WFPC2 Data Reduction Handbook and interpolated Johnson offsets appropriate for 
the spectral 
types of 
XZ~Tau North and South, i.e.\ M2 and M3.5 respectively (Hartigan \& Kenyon 
2003). Our results 
are presented in Table\ \ref{StellarMagnitudes} and the R-band 
WFPC2 data were used to plot a light curve for each component, 
Fig.\ \ref{stellarmagnitudes}. Over the six years of observations, 
XZ~Tau South, although it varies, does so by at most 0.3 magnitudes (in R) 
and has a 
mean R magnitude of around 13.5. In contrast, XZ~Tau North shows an initial 
reduction in brightness of about a magnitude in R until 1998 and thereafter
its brightness increases dramatically by around 3 magnitudes. 
This flaring behaviour means that by 2001 XZ~Tau North was actually the 
brighter star. Similar variations are seen in the I band although the data 
is somewhat more sparse.  

The dramatic brightening of XZ~Tau North suggests it is an EXor. EXors, named 
after their prototype EX Lupi, are extreme classical T Tauri stars that  
periodically undergo outbursts from the UV to the optical. Although increases 
by several magnitudes with rise times of up to a few years have been recorded 
\cite{Herbig89} the changes in these YSOs are not as extreme as in FU Orionis
stars. EXor spectra, for example, even in outburst continue to resemble T Tauri 
stars. The phenomenon is thought to be due to major increases in the underlying 
disk accretion rate, but the number of EXors known is relatively small. The 
proposition that XZ~Tau North is such a YSO is further strengthened
by the HST spectroscopic data of \cite{HartKen03}. As with other EXors the 
spectrum shows not only very strong Balmer lines but also strong Ca II, and 
moderate Na I in emission (see, for example, \cite{parsam02}). 
Finally EXor behaviour would explain why the  
the 1996 HST/FOS acqusition of XZ~Tauri (WG01) unexpectedly locked 
onto the wrong (northern) star. WFPC2 images taken in March 1997, as part of 
HST GO Programme 6725, show XZ~Tauri North comparable in brightness with 
XZ~Tauri South at short optical wavelengths. This also suggests that EXor 
behaviour {\it may be visible at the shortest wavelengths first}. 

\begin{table}

\begin{center}
\scriptsize{
\begin{tabular}{llccc}

\hline
\hline

Filter  &Observation    &Flux Ratio     &XZ South       &XZ North\\
        &Date           &South/North    &Magnitude      &Magnitude\\
\hline

R-band  &1995 Jan 5     &5.11           &13.16          &14.93\\
        &1997 Mar 8     &5.39           &13.47          &15.38\\
        &1998 Mar 25    &9.02           &13.22          &15.68\\
        &1999 Feb 3     &3.66           &13.45          &14.93\\
        &2000 Feb 6     &2.43           &13.32          &14.36\\
        &2001 Feb 10    &0.63           &13.25          &12.82\\
        &               &               &               &\\
I-band  &1995 Jan 5     &4.80           &12.03          &13.78\\
        &1997 Mar 8     &6.30           &11.97          &14.02\\
        &1999 Feb 3     &3.38           &12.04          &13.42\\
        &2000 Feb 6     &1.98           &12.02          &12.82\\ \hline

\end{tabular}}
\end{center}
\small{\caption{Johnson apparent magnitudes of the XZ~Tauri binary. 
Errors in magnitude of $\pm 0.05$ were estimated based on the
affect of changing aperture size, given that the stellar PSFs are
overlapping. The 1995 Jan 5 R-band data is from K97 using Tiny Tim PSF fitting
to the saturated R-band images. The 1997 Mar 8 data is from GO Programme 6725. 
}}
\label{StellarMagnitudes}

\end{table}

\subsection{Modelling the Outflow }

In this section we present the results of a numerical simulation of the
XZ~Tauri outflow in order to check our physical interpretation of the
observations.  The code we use is that described in Downes \& Ray
(1999).  It is a properly upwinded second order (in time and
space), cylindrically symmetric code for solving the inviscid Euler
equations.  In addition to tracking the hydrodynamic variables (density,
velocity and pressure), the code also tracks the ionisation state of
hydrogen, without the assumption of ionisation equilibrium.  It also has
the capacity to track the number density of H$_2$, but this number
density was set to zero for these simulations.

\subsubsection{Initial conditions}

\label{init_conditions}

The gas is taken to be one of solar abundances.  The initial ambient
density is assumed to be 100\,cm$^{-3}$ close to the source, rising to a
value of 600\,cm$^{-3}$ at a distance of $1.25\times10^{16}$\,cm from the
source, (see below for a discussion of why this behaviour is assumed).  The 
ambient pressure was taken to be uniform, giving a temperature of $10^3$\,K 
close to the 
source, and dropping to approximately 160\,K at the rise in
ambient density. The resolution was set to $1\times10^{13}$\,cm in both the 
radial and poloidal (outflow) directions ($r$ and $z$ respectively), and the 
total grid size was $450\times1500$.  To estimate of the mass flux in the 
outflow, we assumed approximately 2\% of the accreted mass ultimately ends 
up in the outflow (\cite{hart95}). A mass accretion rate for XZ~Tau North of 
$1\times10^{-7}$\,$M_{\odot}$\,yr$^{-1}$ was used (Hartigan \& Kenyon 2003) 
based on 
recent photospheric veiling measurements. 

In order to choose parameters for the outflow we must take the following
issues into consideration:
\begin{enumerate}
\item The aspect ratio of the outer bubble/bowshock is small despite the 
fact that the outflow is thought to be highly supersonic with respect 
to the sound speed in the ambient medium,
\item If we assume a constant jet/outflow radius based on the observed inner 
knots, 
i.e.\ $r_{\rm jet}\sim~2\times10^{14}$\,cm, then the inferred mass-flux in the 
outflow would imply a large jet density and specifically one that is much 
higher than the ambient 
density. This cannot be the case for the XZ Tauri outflow, as such a 
highly dense jet would simply plough through the local medium creating a very narrow bowshock, in 
contradiction with our observations. 
\end{enumerate}

One way of explaining the observations is to invoke a moderately collimated 
wind.
A wind with a moderate opening angle will have a density which decreases 
significantly with distance away from the source, and will also have a 
larger radius with distance.  These effects each lead to a broader bowshock, 
in line with the observed morphology of the system.  In addition, if the
flow is not in the plane of the sky, the bowshock will appear to have a
lower aspect ratio due to projection effects.  We assume an angle of 
30\,$\degr$ between the plane of 
the sky and the axis of the outflow.

With these considerations in mind, we set the full opening angle of the 
wind to be 22\,$\degr$ with an initial diameter (full-width 
half-maximum in density) of $4\times10^{14}$\,cm, or 40 grid cells.  
The wind density was given a linear profile across the outflow axis, with 
a density range from $\sim$ 1800\,cm$^{-3}$ along the jet axis to $\sim$ 100\,cm$^{-3}$ 
at the edge, and was chosen so that the total mass-flux was $2\times10^{-9}$\,$M_{\odot}$\,yr$^{-1}$.  Hence we are 
assuming the jet (traced by the knots) does not dominate the dynamics of the system. 

The observed tangential speed of the flow is in the range 
130--200\,km\,s$^{-1}$ 
so we choose a constant wind velocity of 250\,km\,s$^{-1}$, which will give a tangential 
velocity for the jet fluid of roughly 216\,km\,s$^{\rm -1}$. Since the bowshock 
appears to be decelerating significantly (see table \ref{ProperMotions}), 
we need some way to slow it down. Turning off the jet will not, on
its own, give us the observed deceleration since the momentum density of 
the material in the cooled shell between the bowshock and jetshock is very large. 
It was found that the most effective way of decelerating the bowshock was
to impose a significant increase in the ambient density.  As mentioned
above, a six-fold increase was found to roughly reproduce the observed
velocities.

\subsubsection{Simulation results}

\begin{figure}
\begin{center}
{\includegraphics[scale=0.32]{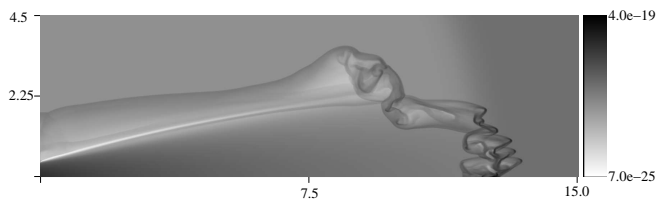}}
\end{center}
\small{\caption{Log-scale plot of the distribution of density at $t=21$\,yrs.  
The distance scales are 
in units of $10^{15}$\,cm and the intensity 
scales are in units of g\,cm$^{-3}$ 
\label{sim_rho}}}
\end{figure}

Figure \ref{sim_rho} contains a greyscale plot of the distribution of density 
at
$t=21$\,yrs, the time at which the simulation roughly matches the 
2001 observations.  The non-zero opening angle of the jet is clearly 
noticeable.  The bowshock itself has a number of irregularities which probably 
arise from the Vishniac instability. Another possible source is the 
Rayleigh-Taylor instability, since the bowshock is decelerating. It can be seen 
that the 
bowshock has begun to encounter the rise in the ambient density (see section 
\ref{init_conditions}). This leads to a marked
decrease in its speed of advance (see table \ref{sim_proper_motions}). The 
behaviour of the 
velocity of the outer shock matches that observed reasonably well (see table 
\ref{sim_proper_motions}).

\begin{table*}
\begin{center}
\scriptsize{\begin{tabular}{llcccc}
\hline\hline
Simulation Age 	&Corresponding 		&Distance	&Bubble 	&Longitudinal 		&Transverse		\\
/Age Interval	&Observation		&from source	&Width		&Speed			&Speed			\\ 
(years) 	&Date/Interval		&(AU)		&(AU)		&(km\,s$^{\rm -1}$)	&(km\,s$^{\rm -1}$)	\\ \hline 
15		&1995 Jan 5 			&590		&288		&-			&-		\\
15 - 18		&1995 Jan 5 to 1998 Mar 25 	&695		&360		&164			&67		\\ 
18 - 19		&1998 Mar 25 to 1999 Feb 3	&722		&386		&148	 		&63		\\ 
19 - 20		&1999 Feb 3 to 2000 Feb 6	&749		&402		&129	 		&51		\\ 
20 - 21		&2000 Feb 6 to 2001 Feb 10	&775		&420		&115	 		&44		\\ \hline
\end{tabular}}
\end{center}
\small{\caption{Projected sizes and speeds of the simulated XZ~Tauri outer 
bowshock as it 
evolves. Distances are quoted for the latter year of the age interval. There is 
a minimum 
error of $\pm 6$\,km\,s$^{\rm -1}$ in the simulation 
speeds. \label{sim_proper_motions}}}
\end{table*}

\begin{figure}
\begin{center}
{\includegraphics[scale=0.28]{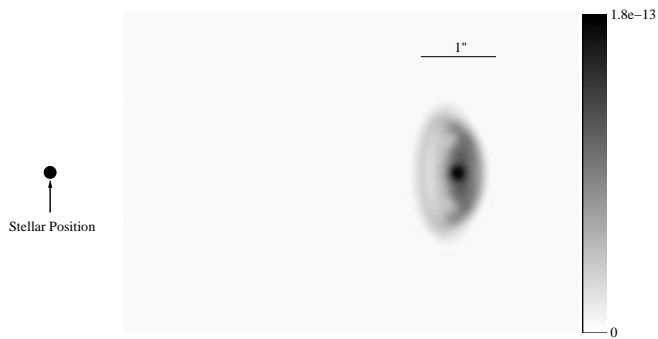}}
\end{center}
\small{\caption{Simulated [SII] image projected onto the sky (assuming
an inclination to the plane of the sky of angle of 30\,\degr) at $t=21$\,yrs 
(see text).  The intensity scales are in units of erg\,s$^{-1}$\,cm$^{-2}$ 
\label{sim_sii}}}
\end{figure}

Figure \ref{sim_sii} is a simulated [SII] image calculated for
the density distribution shown in figure \ref{sim_rho}, assuming an angle to 
the plane of the sky of 30\,\degr. In calculating this we assume that the 
ionisation fractions state of sulphur can be described by coronal ionisation equilibrium (\cite{Arnaud85}) and 
that the line emission is not in Local thermal equilibrium. The emission is 
plotted on a linear scale (similar to that in figure \ref{xztaupm}).  It has been projected onto the sky and 
convolved with a Gaussian of FWHM\,=\,0$\farcs$1.

It can be seen that, broadly speaking, there is reasonable agreement
between the observations and the simulation, at least in terms of the
morphology close to the head of the bowshock.  The ring-like features
arise from the assumption of cylindrical symmetry in the simulation.
The emission would, in fact, be expected to be a little more `clumpy'
rather than ring-like.

Finally, the total [SII] luminosity, calculated from the simulation is 
$1.1\times10^{29}$\,erg\,s$^{-1}$. This is in good agreement with that 
observed (i.e. 4.3~($\pm 3.6$)\,$\times$$10^{\rm 28}$\,erg\,s$^{\rm -1}$, see 
section 
\ref{structure_etc}).  We note however that the bowshock apex appears much 
brighter with respect to the wings than is actually observed.  This is 
most likely due to our use of cylindrical symmetry which results in a focussing of shocked material 
onto the axis in a way which would be very unlikely to happen in three dimensions. 


\section{Conclusions}

Multi-epoch HST/WFPC observations of the XZ~Tau binary and its 
associated outflow have shown considerable changes in the system within only 
6 years, from 1995 to 2001. The presence of {\em two} limb-brightened shock 
fronts is now clearly evident, with a deceleration of 
the outer shock from 146\,km\,s$^{-1}$ to 117\,km\,s$^{-1}$.  Stellar 
photometry 
revealed that the suspected source of the outflow, XZ~Tau North, has flared in 
EXor-type fashion increasing in brightness by 3 magnitudes in R between 1998 
and 2001. Finally, numerical simulations of the outflow produced reasonable
agreement with observation in terms of morphology, dynamical evolution and 
emission line luminosity, using plausible ambient and outflow parameters. 
Deceleration by the amount observed, caused by the ambient medium, should have 
produced a 
much brighter 
bowshock apex than that seen. The cause of this discrepancy is not obvious. 


\vspace {0.3in}
{\bf Acknowledgements} 
\newline

We wish to thank the referee, Dr S. Cabrit, for useful comments and 
suggestions. D.C. and T.P.R. 
would like to acknowledge support for their research from Enterprise Ireland. 
This work was carried 
out as part of the CosmoGrid project, funded under the Programme for Research 
in Third Level 
Institutions (PRTLI) administered by the Irish Higher Education Authority under 
the National
Development Plan and with partial support from the European Regional 
Development Fund.


\end{document}